\documentclass[11pt,a4wide]{article}
\pdfoutput=1
\usepackage{jheppub}
 \usepackage{graphicx}
 \usepackage{bm}
 \usepackage{braket}
\usepackage{epsf,
shadow,pifont}
 \usepackage{amsmath,epsfig}
 \usepackage{amssymb,amsfonts,amsthm,amstext,amscd,dsfont}
\usepackage{latexsym}
\usepackage{slashed}
\usepackage{float}
\usepackage{epsfig}
\usepackage{simplewick}

\relax
\renewcommand{\theequation}{\arabic{section}.\arabic{equation}}

\def\be{\begin{eqnarray}}
\def\ee{\end{eqnarray}}
\def\D{\Delta}
\def\t{\tau}
\def\a{\alpha}
\def\b{\beta}

\def\D{\Delta}
\def\l{\lambda}
\def\g{\gamma}
\def\G{\Gamma}
\def\d{\delta}
\def\nn{\nonumber\\}
\def\pa{\partial}

\def\e{\epsilon}

\newcommand\cO{\mathcal{O}}
\newcommand\<\langle
\renewcommand\>\rangle

\title{Factorized lightcone expansion of conformal blocks}

\author{Wenliang Li}
\affiliation{Okinawa Institute of Science and Technology Graduate University, 1919-1 Tancha, Onna-son, Okinawa 904-0495, Japan}

\emailAdd{lii.wenliang@gmail.com}

\abstract{
We present a factorized decomposition of 4-point scalar conformal blocks near the lightcone, 
which applies to arbitrary intermediate spin and general spacetime dimensions. 
Then we discuss the systematic expansion in large intermediate spin 
and the resummations of the large-spin tails of Regge trajectories. 
The basic integrals for the Lorentzian inversion are given by Wilson functions. 
}

\preprint{}

\begin{document}

\maketitle 
\section{Introduction}
A complex physical process with multi-variables can simplify in extreme situations, 
where the leading behaviour is captured by a product of simpler processes with fewer variables. 
For example, a scattering amplitude factorizes into two disconnected amplitudes with fewer legs 
when an intermediate state is on the mass shell, as the intermediate particle can propagate a large distance. 
\footnote{
Here we consider tree amplitudes. 
The factorization of tree amplitudes also takes place in the soft and collinear limits. 
Similarly, the discontinuities of loop amplitudes are associated with the products of amplitudes with lower loops. }
Based on the simplification by factorization, 
the singular kinematic configurations can furnish the starting points for systematic analytic studies. 

In this paper, we are interested in conformal correlators near the lightcone, 
in the context of the conformal bootstrap \cite{Ferrara:1973yt, Polyakov:1974gs}. 
\footnote{ 
Conformal blocks play a crucial role in the conformal bootstrap program 
and have been studied from the 1970's 
\cite{Ferrara:1973vz, Ferrara:1974nf, Ferrara:1974ny, Dobrev:1975ru, Lang:1992zw}. 
Based on Dolan and Osborn's works \cite{Dolan:2000ut, Dolan:2003hv, Dolan:2011dv}, 
the conformal bootstrap program was revived by the seminal work \cite{Rattazzi:2008pe}. 
We refer to \cite{Poland:2018epd} for a comprehensive review on the numerical conformal bootstrap. }
The lightcone limits are singular. 
When two operators are spacelike separated, causality implies that their commutator vanishes. 
However, timelike-separated operators do not commute.  
The operator ordering is related to the paths of analytic continuation from spacelike to timelike separation. 
The ambiguities in ordering are precisely due to the presence of lightcone singularities at lightlike separation. 
The simplification due to the singular lightcone limits makes analytic studies of the bootstrap equations more tractable. 
In particular, according to the lightcone limit of crossing constraints, the leading behaviour is determined by the vacuum contribution in the cross-channel, 
and the large spin spectrum behaves as mean fields, even at strong coupling 
\cite{Fitzpatrick:2012yx, Komargodski:2012ek}. 
The other cross-channel intermediate states encode the corrections to the generalized free theory. 
Furthermore, the twist spectrum is additive, 
so there are double-twist trajectories and, more generally, multi-twist trajectories. 
\footnote{See \cite{Alday:2007mf} for an earlier discussion of double twists in a more concrete context.} 
One can also study real time dynamics by analytic continuation around the lightcone singularities, 
which also simplifies near the lightcone, such as in the Regge limit. 
\footnote{
The analytic continuation of a conformal block around the lightcone limits leads to 
a linear combination of the independent solutions of the Casimir equation. 
The different solutions have the same Casimir eigenvalues, 
and are related by Weyl reflections, such as the shadow transform $\D\rightarrow d-\D$. }
The real-time dynamics is central to many analytic developments in recent years, 
such as the chaos bound \cite{Maldacena:2015waa} and 
a proof of the average null energy condition \cite{Hartman:2016lgu}. 

The lightcone bootstrap originally gives rise to asymptotic results at large spin 
\cite{Kaviraj:2015cxa,Alday:2015eya,Kaviraj:2015xsa,Alday:2015ota,Alday:2015ewa,Alday:2016mxe,Alday:2016njk,Alday:2016jfr,Simmons-Duffin:2016wlq,Dey:2017fab,Cardona:2018dov}.  
In \cite{Alday:2015ota,Simmons-Duffin:2016wlq}, it was shown that 
the asymptotic methods are consistent with the numerical results down to spin two for the 3d Ising CFT. 
By imposing good Regge behaviour, the Lorentzian inversion formula proposed by Caron-Huot 
reconstructs the OPE data from the double discontinuities of the conformal correlators \cite{Caron-Huot:2017vep,Simmons-Duffin:2017nub,Kravchuk:2018htv}, 
which also establishes analyticity in spin assumed in the conformal Regge theory \cite{Costa:2012cb}. 
These double discontinuities encode the essential information about lightcone singularities 
in terms of double commutators at timelike separation. 
Based on the Lorentzian inversion formula, 
the lightcone bootstrap becomes a method that gives reliable results at finite spin 
\cite{Liu:2018jhs,Cardona:2018qrt, Albayrak:2019gnz, Li:2019dix, Albayrak:2020rxh, Liu:2020tpf,Caron-Huot:2020ouj}, 
as the nonperturbative contributions in the large-spin expansion are also captured. 
More recently, it was proposed in \cite{Caron-Huot:2020adz} that 
the CFT dispersive sum rules \cite{Caron-Huot:2020adz,Carmi:2019cub,Mazac:2019shk,Penedones:2019tng,Carmi:2020ekr} 
can provide a more rigorous alternative to the above lightcone bootstrap. 

We will focus on the 4-point scalar conformal blocks,
which play a crucial role in the analytic bootstrap studies based on 4-point scalar correlators
\be
\langle \phi_1\,\phi_2\,\phi_3\,\phi_4\rangle
=
\bigg(\frac {x_{24}}{x_{14}}\bigg)^{2a}\,
\bigg(\frac {x_{14}}{x_{13}}\bigg)^{2b}\,
\frac {\mathcal G(u,v)} {x_{12}^{\D_1+\D_2}\, x_{34}^{\D_3+\D_4}}\,.
\label{4-p-fn}
\ee 
Here $x_i$ indicates the position of the external primary operators $\phi_i$, 
and $x_{ij}=|x_i-x_j|$ denotes the distance between $\phi_i$ and $\phi_j$. 
The external operators can have different scaling dimensions, 
and their differences are encoded in
\be
a=\frac {\D_1-\D_2} 2\,,\quad
b=\frac{\D_3-\D_4} 2\,.
\ee
The 4-point function is determined by conformal symmetry up to a function of 
two cross-ratios
\be
u=\frac {x_{12}^2\, x_{34}^2}  {x_{13}^2\, x_{24}^2},\qquad
v=\frac {x_{14}^2\, x_{23}^2}  {x_{13}^2\, x_{24}^2}\,,
\ee
as they are invariant under conformal transformations.  
In this work, we will mainly use  the $(z,\bar z)$ variables, which are related to $(u,v)$ by 
\be
u=z\bar z\,,\quad
v=(1-z)(1-\bar z)\,.
\ee
In the Lorentzian signature, $(z,\bar z)$ are real, independent coordinates. 
Although the correlator is symmetric in $(z,\bar z)$, we will break this symmetry and focus on the regime $0\leq z<\bar z\leq 1$. 
The $z\rightarrow 0$ limit is the direct-channel lightcone limit, as a pair of operators in the direct-channel OPE becomes lightlike separated. 
Similarly, the $\bar z \rightarrow 1$ limit corresponds to the cross-channel lightcone limit. 

The correlator $\mathcal G(u,v)$ can be decomposed into conformal blocks labelled by primary operators
\be
\mathcal G(u,v)=\sum_{\cO_i} P_i\, G_{\t_i,\ell_i}^{(d,a,b)}(z,\bar z), 
\ee
where $\t_i,\ell_i$ are the twist and spin of $\cO_i$. 
The contributions of descendants are encoded in the conformal blocks 
according to their primary operators. 
These 4-point scalar conformal blocks satisfy a differential equation associated with the quadratic Casimir \cite{Dolan:2003hv}
\be
\mathcal D^{(2)}\, G^{(d,a,b)}_{\t,\ell}(z,\bar z)
=\frac 1 2 \big[(\t+\ell)(\t+\ell-d)+\ell(\ell+d-2)\big]\,
G^{(d,a,b)}_{\t,\ell}(z,\bar z)\,,
\label{Casimir-eq}
\ee
where  the quadratic conformal Casimir and $SL(2,\mathbb R)$ Casimir are
\be
\mathcal D^{(2)}=\mathcal D_z+\mathcal D_{\bar z}
+(d-2)\frac{z\bar z}{z-\bar z}
\big[(1-z)\pa_z-(1-\bar z)\pa_{\bar z}\big]\,,
\ee
\be
\mathcal D_x=(1-x)\,x^2\pa_x^2-(1-a+b)\,x^2\,\pa_x+ab\,x\,.
\ee
We will also use the $SL(2,\mathbb R)$ block
\be
k^{(a,b)}_{2h}(x)=x^{h}
{}_2F_1(
h-a,h+b,
2h
;\,x)\,,
\ee
which solves the $SL(2,\mathbb R)$ Casimir equation 
\be
\mathcal D_x\, k^{(a,b)}_{2h}(x)=h(h-1)\, k^{(a,b)}_{2h}(x)\,.
\label{SL2R-Casimir-eq}
\ee
To simplify the notation, we will not write ${(a,b)}$ explicitly in some equations. 

\subsection{Factorization at large spin}
It was shown in \cite{Fitzpatrick:2012yx} that a conformal block factorizes in the large spin and crossed lightcone limit
\be
G^{(d,a,b)}_{\t,\ell}(z, \bar z )=k^{(a,b)}_{\t+2\ell}(\bar z) \,F^{(d,a,b)}(\t, z)+\dots\,,
\label{large-spin-factorization}
\ee 
where the first part depends on the conformal spin $\b=\t+2\ell$ and cross-ratio $\bar z$, and the second part is a function of the twist $\t$ and cross-ratio $z$. 
The first part is explicitly given by the $SL(2,\mathbb R)$ block associated with the symmetry of the direct-channel lightcone, 
which is independent of the spacetime dimension $d$.   
On the other hand, the second part is related to the cross-channel lightcone, and explicitly depends on the spacetime dimension $d$. 
Recently, this factorization was used to study the cross-channel contributions of large spin operators in the 3d Ising CFT \cite{Caron-Huot:2020ouj}. 
However, for the cross-channel contributions at low spin, one cannot use this simple large-spin formula. 
\footnote{In \cite{Caron-Huot:2020ouj}, the low spin contributions were computed using the dimensional reduction \cite{Hogervorst:2016hal}, 
in which 3d conformal blocks are approximated by 
2d conformal blocks. } 
It will be useful to have a factorization formula for arbitrary spin, so one can apply it to low spin as well. 
In addition, one can use the general-spin formula to derive the subleading terms in the large-spin expansion, 
which is necessary for a more precise approximation of the large spin contributions. 
\footnote{In fact, \cite{Fitzpatrick:2012yx} only derived the explicit expression of $F^{(d,a,b)}(\t, z)$ in 2d and 4d, as the conformal blocks are known in simple closed form. 
We will give the general expression of $F^{(d,a,b)}(\t,z)$ in \eqref{large-spin-expansion-explicit}, 
which is also related to the $SL(2,\mathbb R)$ block. }

In \cite{Li:2019dix}, a general formula for the crossed lightcone limit of 4-point scalar conformal blocks was found. 
The explicit expression can be neatly packaged into a two-variable hypergeometric function, 
which will be given later in \eqref{cross-lightcone} and discussed in more detail. 
For physical spin $\ell$, it reduces to a sum of $(\ell+1)$ Gaussian hypergeometric functions. 
Therefore, to obtain a finite-spin extension of \eqref{large-spin-factorization}, 
we just need to consider the product of the analytic expressions for the two lightcone limits, 
i.e. the $SL(2,\mathbb R)$ block for the direct-channel and the more recent formula \eqref{cross-lightcone} for the cross-channel. 
\footnote{Note that the small $(1-\bar z)$ expansion of $k_{\t+2\ell}(\bar z)$ gives rise to two ${}_2F_1$ functions, 
which are related by interchanging external dimensions. 
They should be multiplied by the associated cross-channel formulae. }
For arbitrary spin, there is no large-spin suppression, 
so the product of the two lightcone blocks only captures the leading behaviour near the lightcone. 
This is sufficient for the study of leading twist trajectories. 
However, for more systematic investigations, we need to know the precise subleading terms in the lightcone expansion. 

\subsection{Lightcone expansions to all orders}
For the order-by-order lightcone expansion in one cross-ratio, it is more natural to use $(u,v)$, instead of $(z,\bar z)$.  
This can already be seen from the compact formula for the conformal blocks of spin-0 intermediate operators \cite{Dolan:2000ut},  
whose explicit expression can be found in \eqref{spin-0-G}. 
According to the spin-0 formula, it is clear that the small $u$ or $v$ expansion leads to simpler results than the case of $z$ or $(1-\bar z)$. 
At each order of the single lightcone expansions, 
the dependence on the other cross-ratio is encoded in one ${}_2F_1$ function with simple expansion coefficients. 

For arbitrary spin, the explicit formulae for single lightcone expansions were recently obtained in \cite{Li:2019cwm}. 
The dependence on the other cross-ratio is encoded in $(\ell+1)$ Gaussian hypergeometric functions for physical spin $\ell$. 
Let us discuss the structure of the expansion coefficients. 
The general formula for the cross-channel expansion coefficients is given by multiple summations. 
One of them involves shifted series coefficients of $SL(2,\mathbb R)$ blocks, 
so is closely related to the direct-channel expansion. 
On the other hand, if we examine the direct-channel expansion coefficients, 
we can notice that they involve shifted parameters of cross-channel basis functions.  
So there should exist a unifying decomposition for the direct and crossed lightcone expansions.  

Then it is natural to consider the factorized lightcone expansion in terms of the products of two basis functions. 
If some complexity of the single lightcone expansions is transferred to the additional basis functions, 
then the expansion coefficients should take a simpler form. 
Inspired by the large-spin factorization, we switch to the $(z,\bar z)$ variables. 
It turns out that this is indeed the case. The resulting decomposition is remarkably simple! 
We will present the explicit formula for the factorized lightcone expansion in \eqref{factorized-decomposition}-\eqref{basis-g}.

\section{Even spacetime dimensions are special}
As another source of inspiration for the factorized decomposition, 
it is known that conformal blocks in even dimensions can be decomposed into a finite sum in factorized form in terms of $(z,\bar z)$. 
\footnote{To be more precise, the inseparable part is encoded in simple integer powers of $(z-\bar z)$.} 
The explicit expressions in $d=2,4,6$ were discovered by Dolan and Osborn \cite{Dolan:2000ut, Dolan:2003hv}. 
One may wonder if conformal blocks in general dimensions can also be decomposed into a finite number of factorized summands. 
In this section, we want to argue that this is not likely for generic scaling dimensions and intermediate spin. 
\footnote{Conformal blocks in general dimensions can become simpler for special scaling dimensions, i.e. when $\D, a, b$ take special values. 
The important examples are intermediate operators that saturate the unitarity bounds, such as the stress tensor, 
where some descendants become null states due to the conserved current equations. 
Below the unitarity bound, there can also be partially conserved currents, which are non-unitary representations of the conformal algebra. }
We will also discuss some previous results that 
lead us to the factorized lightcone expansion for general $d$ in \eqref{factorized-decomposition}. 

\subsection*{Conformal blocks with $\ell=0$}
Let us consider the simplest case of $\ell=0$. 
In general dimensions, the conformal block for an intermediate operator of spin-0 reads \cite{Ferrara:1973vz,Ferrara:1974nf}
\be
G^{(d,a,b)}_{\D,0}(u,v)=v^{\frac {a-b}2}
\Bigg(\frac {\G(b-a)\,v^{\frac {a-b}2}u^{\D/2}}{(\D/2)_{-a}(\D/2)_b}
F_4\bigg[
\begin{matrix}
\D/2+a, \D/2-b\\
\D-\frac {d-2} 2, 1+a-b
\end{matrix};u,v
\bigg]+(a\leftrightarrow b)\Bigg)\,,\quad
\label{spin-0-G}
\ee
where Appell's $F_4$ function is a two-variable hypergeometric function
\be
F_4\bigg[
\begin{matrix}
A_1, A_2\\
B_1,B_2
\end{matrix}; u,v
\bigg]
=\sum_{m,n=0}^\infty
\frac{(A_1)_{m+n}\,(A_2)_{m+n}}{(B_1)_{m}\,(B_2)_{n}}\frac{u^m\,v^n}{m!\,n!}\,.
\ee
Note that $(x)_y=\G(x+y)/\G(x)$ and $\G(x)$ is the Gamma function. 
In general, the $F_4$ function is absolutely convergent for $|u|^{1/2}+|v|^{1/2}<1$, 
which contains the Lorentzian regime $0\le z<\bar z\leq 1$ under consideration. 

When $B_1+B_2=1+A_1+A_2$, it was shown by Bailey that the $F_4$ function factorizes into a product of two ${}_2F_1$ functions \cite{Bailey}. 
This is precisely the case of 2d conformal blocks \cite{Ferrara:1974nf}. 
For $d=4$, conformal blocks also admit finite decompositions, 
and the corresponding $F_4$ functions can be decomposed into two terms.  
The precise decomposition formulae for these $F_4$ functions are given in \eqref{F4-1-balanced} and \eqref{F4-0-balanced}.  
To our knowledge, a generic $F_4$ function cannot be written as a finite sum in factorized form. 
The above $d=2,4$ conformal blocks are special in that the $F_4$ functions are 0-balanced or 1-balanced. 
More generally, the even-dimensional conformal blocks for intermediate scalars are associated with 
integer-balanced $F_4$ functions of $(u,v)$, 
and this property is independent of the external and intermediate scaling dimensions. 
Therefore, only in even dimensions, 
the factorized decomposition of generic conformal blocks can be written as a finite sum. 

Nevertheless, Burchnall and Chaundy found that a generic $F_4$ function admits 
a nice factorized decomposition (see eq.(54) in \cite{Burchnall-Chaundy}).  
The related decomposition formulae are recorded in \eqref{F4-decomposition} and \eqref{F4-derivative}. 
As a result, we can express an $\ell=0$ conformal block 
as an infinite sum of factorized summands in $(z,\bar z)$. 
The explicit $F_4$ function in \eqref{spin-0-G} becomes
\be
&&F_4\bigg[
\begin{matrix}
\D/2+a, \D/2-b\\
\D-\frac {d-2} 2, 1+a-b
\end{matrix};u,v
\bigg]
\nn&=&
\sum_{k=0}^\infty \frac{\big(\frac {d-2} 2\big)_k}{k!}\,
\frac{(\D/2+a)_k\,(\D/2-b)_k}
{(\D-\frac {d-2} 2)_k\,(1+a-b)_k}\,z^k
{}_2F_1\bigg[
\begin{matrix}
\D/2+a+k, \D/2-b+k\\
\D-\frac {d-2} 2+k
\end{matrix};z
\bigg]
\nn&&\qquad\times
(1-\bar z)^k\,
{}_2F_1\bigg[
\begin{matrix}
\D/2+a+k, \D/2-b+k\\
1+a-b+k
\end{matrix};1-\bar z
\bigg]
\,,
\label{spin-0-decomposition}
\ee
where the parameters in the expansion coefficients and ${}_2F_1$ functions
are directly related to the $F_4$ parameters. 
This nice factorized decomposition for $\ell=0$ motivates us to consider the generalization to arbitrary spin $\ell$. 

Note that $(d-2)/2$ is the difference between the sums of $F_4$ parameters in the top and bottom rows.
For $d=2$, we can see explicitly that 
the $k$-summation terminates at $k=0$, and the factorized decomposition contains only one term. 
For generic $d$, we do not expect that the factorized decompositions terminate at finite order when scaling dimensions are generic. 

\subsection*{Cross-channel lightcone limit}
The simplicity of even-dimensional conformal blocks can also be noticed from the cross-channel lightcone limit. 
\footnote{
The direct-channel lightcone limit does not depend on $d$, 
and always takes a compact form
\be
G^{(d,a,b)}_{\t,\ell}(z,\bar z)\big|_{z\rightarrow 0}=z^{\t/2}
k^{(a,b)}_{\t+2\ell}(\bar z)\,.
\ee}
The closed-form expression reads \cite{Li:2019dix} 
\be
G^{(d,a,b)}_{\t,\ell}(z,\bar z)\Big|_{\bar z\rightarrow 1}&=&
v^{\frac {a-b} 2}
\Bigg(
\frac{\G(b-a)\,v^{\frac{a-b}2}}
{( \b/2)_{-a}\,(\b/2)_b}\,\frac {z^{\t/2}} {(1-z)^{\frac {d-2}2+a-b}}
\nn&&\times\,
F^{0,2,2}_{0,2,1}
  \bigg[
\Big|
\begin{matrix}
 -\ell,\, 3-d-\ell \\
2- d/ 2 -\ell\,,\g
\end{matrix}\,
\Big|
\begin{matrix}
  \g/2-a , \, \g/ 2+b\\
 \b/2+\g/ 2
\end{matrix}\,\Big| 
z,-z
\bigg]
+ \,(a\leftrightarrow b)
\Bigg)\,,\qquad
\label{cross-lightcone}
\ee 
where $\b=\t+2\ell$, $\g=\t-d+2$, and $F^{0,2,2}_{0,2,1}$ is a Kamp\'e de F\'eriet function. 
Note that the crossed lightcone limit satisfies a quartic differential equation \cite{Caron-Huot:2017vep}. 
More explicitly, the Kamp\'e de F\'eriet function in \eqref{cross-lightcone} can be decomposed into
 \be
 \frac {z^{\t/2}} {(1-z)^{\frac {d-2}2+a-b}}
\,F^{0,2,2}_{0,2,1}[\dots]
 =\sum_{n=0}^\infty\, C_{0,0,n}\, g_{0,n}(z)\,.
 \label{KDF-expansion}
 \ee
 The expansion coefficients are
 \be
C_{0,0,n}=\frac{(-1)^n}{n!}\,\frac{(-\ell)_n\,(3-d-\ell)_n}{(2-d/2-\ell)_n}
\frac{(\g/2-a)_n\,(\g/2+b)_n}{(\g)_n\,(\b/2+\g/2)_n}\,,
\label{C00n}
\ee
so the $n$-summation terminates when $\ell$ is a nonnegative integer. 
The basis functions are
 \be
\quad g_{0,n}(z)=\frac{z^{\t/2+n}}{(1-z)^{\frac{d-2}2+a-b}}\,
{}_2F_1\bigg[
\begin{matrix} 
\g/2-a+n,\g/2+b+n \\
\b/2+\g/2+n
\end{matrix}
;\,z\bigg]\,.
\label{g0n}
\ee
The lower index $0$ implies that they are the zeroth order expression in the small $(1-\bar z)$ expansion. 
These expansion coefficients and basis functions will be generalized to higher orders in Section \ref{Factorized-lightcone-expansion}. 
For $d=2,4$, the coefficients $C_{0,0,n}$ are simpler
because $(2-d/2-\ell)_n$ cancel out a Pochhammer symbol in the numerator. 
Then the Kamp\'e de F\'eriet function is independent of $\ell$ and reduces to a Gaussian hypergeometric function 
\be
F^{0,2,2}_{0,2,1}
[\dots]\big|_{d=2,4}=
{}_2F_1\big(
\g/2-a,\g/2+b,
\g
;\,z\big)\,,
\label{t-2d-4d}
\ee
which is associated with an $SL(2,\mathbb R)$ block parameterized by $\g=\t-d+2$. 
Furthermore, the $d=2, 4$ conformal blocks have simple closed-form expressions
\be
G^{(d,a,b)}_{\t,\ell}(z,\bar z)\Big|_{d=2,4}
=\frac{\G(\b/2)^2}{\G(\b)}\Big(\frac{z\bar z}{\bar z-z}\Big)^{\frac{d-2}2}\,k_{\g}(z)\,k_{\b}(\bar z)+(z\leftrightarrow \bar z)\,,
\ee
where the second term is required by the symmetry in $(z,\bar z)$ and becomes redundant when $d=2,\ell=0$. 
\footnote{One should be careful about the order of the even-$d$ and integer-$\ell$ limits. In this sense, the even-$d$ limits are singular, 
so one can achieve the special simplifications. 
There might be an interesting interplay between $d/2$ and $\ell$ such that special simplification can take place. } 
In Section \ref{Large-spin-expansion}, 
we will revisit the factorized lightcone expansion at large spin.  
We will show that the leading term at large spin is $d$-independent up to $\big(\frac z {1-z}\big)^{(d-2)/2}$, 
so the $SL(2,\mathbb R)$ block $k_{\g}(z)$ also appears there. 

For higher even $d$, the coefficients $C_{0,0,n}$ are also simpler than the general case. 
In fact, there exists an alternative formula for the cross-channel lightcone limit, 
which manifestly reduces to a finite sum of ${}_2F_1$ functions for even $d$. 
The explicit formula is given in \eqref{t-lightcone-even-d}.  
According to this equivalent expression, the cross-channel lightcone limit reduces to  
at most $(1+|d-3|)/2$ independent ${}_2F_1$ functions in even dimensions, 
so the $d=2$ and $d=4$ decompositions have only one term. 
Note that the spin $\ell$ is generic here. 
The cases of higher even dimensions also admit finite decompositions, but the complexity of the exact expression grows with $d$, 
as expected from the generic $d$ formula. 
\footnote{As in \cite{Fitzpatrick:2013sya}, 
one may consider an expansion in large $d$, which significantly simplifies the analytic expression of conformal blocks. 
This is similar to the expansion in large $\ell$. }

\section{Factorized lightcone expansion of conformal blocks}
\label{Factorized-lightcone-expansion}
To generalize \eqref{spin-0-decomposition} and \eqref{KDF-expansion}, 
we study the factorized lightcone expansion of generic 4-point scalar conformal blocks. 
A remarkably simple formula is obtained:
\be
G^{(d,a,b)}_{\t,\ell}(z,\bar z)
=v^{\frac {a-b}2}
\Bigg(\frac {\G(b-a)\,v^{\frac {a-b}2}}{(\b/2)_{-a}(\b/2)_b}
\sum_{k=0}^\infty \sum_{m,n}\,
C_{k,m,n}\,
f_{k,m}(1-\bar z)\,g_{k,n}(z)
+(a\leftrightarrow b)
\Bigg)\,,\quad
\label{factorized-decomposition}
\ee
where $k$ denotes the order of the lightcone expansion and $v=(1-z)(1-\bar z)$. 
It is convenient to use the 
anomalous dimension $\g$
and conformal spin $\b$:
\be
\g=\t-d+2\,,\quad
\b=\t+2\ell\,,
\ee
where twist is defined as $\t=\D-\ell$. 
\footnote{To be precise, $\g$ is the anomalous dimension for spinning operators. }
The expansion coefficients also factorize:
\be
C_{k,m,n}
&=&
\frac{(-1)^{n}\,(-k)_m\,(-\ell)_{n}\,(\frac{d-2} 2-m)_{k+m}}
{k!\,m!\,n!}\,
\frac{(\t+\ell-1)_{k-m}\,(3-d-\ell-k+m)_{n}}
{(\g)_{n}\,(2-d/2-\ell)_{m+n}}
\nn&&\times
\frac{(-\b/2+a+1)_m\,(-\b/2-b+1)_m}
{(1+a-b)_m}\,
\frac{(\g/2-a)_{n}\,(\g/2+b)_{n}}
{(\b/2+\g/2)_{k+n}}
\,,
\label{C-kmn}
\ee
where the essential terms are in the first line and they are independent of $(a,b)$. 
The second line of \eqref{C-kmn} is closely related to the basis functions,  
and can be absorbed into their definitions in the form of hypergeometric series. 
The basis functions are associated with ${}_2F_1$ functions
\be
f_{k,m}(1-\bar z)=
\frac{(1-\bar z)^k}{\bar z^{\b/2-1+k-m}}\,{}_2F_1
\bigg[
\begin{matrix}
-\b/2+a+1+m,-\b/2-b+1+m\\
1+a-b+m
\end{matrix}; 1-\bar z
\bigg]\,,
\label{basis-f}
\ee
\be
g_{k,n}( z)=
\frac{z^{\t/2+k+n}}{(1-z)^{\frac {d-2}2+a-b+k}}\,
{}_2F_1
\bigg[
\begin{matrix}
\g/2-a+n, \g/2+b+n\\
\b/2+\g/2+k+n
\end{matrix}; z
\bigg]\,,
\label{basis-g}
\ee
which can also be expressed in terms of $y=z/(1-z)$ and $\bar y=(1-\bar z)/\bar z$. 
To approximate a conformal block near the direct or crossed channel lightcone, 
we can truncate the summation to $k= k_\text{cutoff}$, 
which neglects some terms of higher order than $z^{\t/2+k_\text{cutoff}}$ or $(1-\bar z)^{k_\text{cutoff}}$. 
The truncated sum still contains an infinite number of higher order terms, 
so gives a better approximation than the order-by-order expansion in one variable. 
At large spin, the low order terms in $k$ capture the dominant contributions, 
and hence the factorized decomposition \eqref{factorized-decomposition} is particularly suitable for the systematic expansion in large spin, 
which will be discussed in Section \ref{Large-spin-expansion}. 
The decomposition coefficients are nonzero when
\be
0\le m\le k\,,
\ee
so the $\bar z$-dependence is encoded in at most $(k+1)$-independent ${}_2F_1$ functions at order $k$. 
\footnote{For the order-by-order expansion in small $z$, 
the number of independent basis functions for the $\bar z$ dependence increases as $2k+1$. } 

For physical spin $\ell=0,1,2,\dotsb$, 
the $n$-summation terminates at $n=\ell$, then we have
\be
0\le n\le\ell\,. 
\ee 
Furthermore, the $m$-summation of $f_{k,m}$ reduces to at most $(\ell+1)$ Gaussian hypergeometric functions.
This becomes manifest if we use a different set of basis functions
\be
\tilde f_{k,m}(1-\bar z)=\bar z^{\b/2-m}\,(1-\bar z)^k\,{}_2F_1
\bigg[
\begin{matrix}
\b/2+a+k-m, \b/2-b+k-m\\
1+a-b+k-m
\end{matrix}; 1-\bar z
\bigg]\,.
\ee
Accordingly, we have
\be
G^{(d,a,b)}_{\t,\ell}(z,\bar z)
=v^{\frac {a-b}2}
\Bigg(\frac {\G(b-a)\,v^{\frac {a-b}2}}{(\b/2)_{-a}(\b/2)_b}
\sum_{k=0}^\infty \sum_{m,n=0}^\ell
\tilde C_{k,m,n}\,
\tilde f_{k,m}(1-\bar z)\,g_{k,n}(z)
+(a\leftrightarrow b)
\Bigg)\,,\quad
\label{factorized-decomposition-tilde}
\ee
where the expansion coefficients can be expressed in terms of \eqref{C-kmn}
\be
\tilde C_{k,m,n}&=&\sum_{m'=0}^m\,\frac{(k-m+m')!}{(k-m)!\,(m')!}
\frac{
(\b/2+a)_{k-m}\,(\b/2-b)_{k-m}\,(1-\b)_{m'}}
{(-\b/2+a+1)_{k-m+m'}\,(-\b/2-b+1)_{k-m+m'}}
\nn&&\qquad\times
(k-m+1+a-b)_{m'}\,C_{k,k-m+m',n}
\,.\quad
\ee
Therefore, the concrete expression of the factorized decomposition further simplifies at low spin. 
The complexity increases with spin, 
but there is an emergent simplicity at large spin as mentioned earlier. 
In Section \ref{Large-spin-expansion}, we will discuss the systematic large-spin expansion, 
as an application of the factorized decomposition \eqref{factorized-decomposition}. 

In the limit $a-b\rightarrow 0$, 
the general formula \eqref{factorized-decomposition} becomes divergent due to the Gamma functions $\G(a-b)$ and $\G(b-a)$, 
but the divergent contributions cancel out. 
The finite pieces are given by a regular part and a logarithmic part, 
arising from the derivative with respect to $(a-b)$. 
The explicit expression is
\be
G^{(d,a,a)}_{\t,\ell}(z,\bar z)
&=&
\frac {(-1)}{(\b/2)_{-a}(\b/2)_{a}}
\Big(\log v+H_{\b/2+a-1}+H_{\b/2-a-1}+\pa_a-\pa_b\Big)
\nn&&\qquad\qquad\times
\sum_{k=0}^\infty \sum_{m,n}\,
C_{k,m,n}\,f_{k,m}(1-\bar z)\,g_{k,n}(z)\Big|_{b\rightarrow a}
\,.\quad
\label{degenerate}
\ee 
When $a=b=0$, an especially interesting case is the conformal blocks with intermediate conserved currents, 
which can be obtained by taking the limit $\g\rightarrow 0$. 
For the spin-2 conserved current, i.e. the stress tensor 
\footnote{The multi-stress tensor operators can also be studied by the Lorentzian inversion \cite{Li:2019zba,Li:2020dqm}.}, 
the $\log v$ part is consistent with the recent results in \cite{Li:2020dqm}. 

Let us explain how we arrive at the complete decomposition formula \eqref{factorized-decomposition}. 
We use the general formula of the cross-channel lightcone expansion in \cite{Li:2019cwm} 
to derive the double expansion in small $z$ and small $(1-\bar z)$. 
According to the results in \cite{Li:2019cwm}, it is expected that for physical spin $\ell$ 
we need $(\ell+1)$ basis functions to encode the $z$-dependence at each order. 
The basis functions $g_{k,n}(z)$ can be inferred from 
the spin-0 decomposition \eqref{spin-0-decomposition} and the $k=0$ basis functions \eqref{g0n}. 
For the $\bar z$ dependence, we know the precise zeroth-order basis function $ f_{0,0}(\bar z)$ from 
the direct-channel lightcone limit, i.e. the $SL(2,\mathbb R)$ block. 
To find natural generalizations of $f_{0,0}(\bar z)$ at higher order, 
we subtract the double series in $(1-\bar z, z)$ by the zeroth order contribution  
$\sum_n C_{0,0,n}\, f_{0,0}(1-\bar z)\,g_{0,n}(z)$, 
and compute the first order series coefficients in the small $z$ expansion. 
The general result is associated with a sum of two ${}_2F_1$ functions, 
suggesting two basis functions $f_{k,m}(1-\bar z)$ at order $k=1$, 
but the choices are not unique. 
Then we also compute the expansion coefficients of $g_{1,n}(z)$, 
and the general expression can be decomposed into two terms again. 
As we are looking for a factorized decomposition, the coefficients of $g_{1,n}(z)$ should 
be equal to the small $(1-\bar z)$ limit of 
$\sum_{m=0,1} C_{1,m,n}\,f_{1,m}(1-\bar z)$. 
By matching the two sides, we find that \eqref{basis-f} provides natural basis functions 
because the resulting $C_{1,0,n}, C_{1,1,n}$ also factorize, as in the $k=0$ case in \eqref{C00n}. 
Based on these basis functions, we further compute the expansion coefficients at order $k=2$,  
and find the general form of $C_{k,m,n}$ up to normalization. 
Then we compute the $k=3$ coefficients and determine the $(k,m)$-dependent normalization.  
In this way, we obtain the complete decomposition formula \eqref{factorized-decomposition},  
together with \eqref{C-kmn}, \eqref{basis-f}, \eqref{basis-g}.  
Using the Casimir equation \eqref{Casimir-eq}, we test the factorized decomposition to order $k=10$, 
which can be carried out efficiently by setting the parameters to rational numbers.  

In the lightcone bootstrap, we need to compute the Lorentzian inversion of cross-channel conformal blocks
\be
&&\frac{u^{(\D_1+\D_2)/2}}{v^{(\D_2+\D_3)/2}}\,
G_{\t',\ell'}^{(d,a',b')}(1-\bar z,1-z)
=\sum_{k,m,n}
\frac{(z\bar z)^{(\D_1+\D_2)/2}(z\bar z)^{\frac {a'-b'}2}}
{[(1-z)(1-\bar z)]^{(\D_2+\D_3)/2}}
\nn&&\qquad\qquad\times\left(\frac {\G(b'-a')\,(z\bar z)^{\frac {a'-b'}2}}{(\b'/2)_{-a'}(\b'/2)_{b'}}\,
 C^{(a',b')}_{k,m,n}\,
f^{(a',b')}_{k,m}(z)\,
g^{(a',b')}_{k,n}(1-\bar z)
+(a'\leftrightarrow b')
\right)\,, \qquad
\ee 
which involves a typical $\bar z$ integral. 
We use the primes to emphasize that the parameters are for the cross-channel, such as
\be
a'=\frac{\D_3-\D_2}2\,,\quad
b'=\frac{\D_1-\D_4}2\,. 
\ee 
The general result of the basic $\bar z$ integral reads
\be
&&\int_0^1 \frac{d\bar z}{\bar z^2}\,
k^{(-a,-b)}_\b(\bar z)\,
\frac{\bar z^{(\D_1+\D_2)/2}}{(1-\bar z)^{(\D_2+\D_3)/2}}\,\bar z^{a'-b'}
\,g^{(a',b')}_{k,n}(1-\bar z)
\nn&=&\G(\b)\,
\G\Big(\frac{\g'+\b'}2+k+n\Big)\,
\G\Big(\frac{\t'-\D_2-\D_3}2+k+n+1\Big)\,
\G\Big(\frac{\t'-\D_1-\D_4}2+k+n+1\Big)
\nn&&\times\,
\G\Big(\frac{\b+\g'-\t'+\D_1+\D_2}2-k-1\Big)\,
\G\Big(\frac{\b+\b'-\t'+\D_3+\D_4}2-n-1\Big)
\nn&&\times\,
\psi\Big(\frac{\b+\b'+\g'-\D_1+\D_2}2+k+n-1;\,\frac{\b-\D_1+\D_2}2,
\frac{\b'-\D_1+\D_4}2+k,
\nn&&\qquad
\frac{\g'+\D_2-\D_3}2+n,
\frac{\t'-\D_1-\D_4}2+k+n+1,
\frac{\b'+\g'-\t'+\D_2+\D_3}2-1
\Big)\,.\qquad\quad
\label{g-integral}
\ee
We have introduced a very well-poised ${}_7F_6$ function
\be
\psi(A;B_1,B_2,B_3,B_4,B_5)
=\frac{\G(A+1)\,{}_7F_6\Bigg[
\begin{matrix}
A, A/2+1, B_1,\dots, B_5
\\
A/2, A-B_1+1,\dots, A-B_5+1
\end{matrix}; \,1
\Bigg]}{\G\big(2A+2-\sum_{k=1}^5\,B_k\big)\prod_{k=1}^5\,\G(A-B_k+1)}\,,\qquad
\ee
which is also known as the Wilson function \cite{Wilson, Groenevelt}. 
The ${}_7F_6$ function can be written as a sum of two 1-balanced ${}_4F_3$ functions.  
The Wilson function has several equivalent expressions due to its symmetry. 
The other part with $(a'\leftrightarrow b')$ can be obtained by interchanging $(\D_1,\D_2)$ and $(\D_3,\D_4)$, 
due to a symmetry of the direct-channel OPE. 
To derive \eqref{g-integral}, it is useful to consider the variable $\bar y=(1-\bar z)/\bar z$ and Mellin-Barnes integral \cite{Liu:2018jhs}. 

\section{Application to the large-spin expansion}
\label{Large-spin-expansion}
In the lightcone bootstrap, a subtle point is that the conformal-block summations and the lightcone limits may not commute. 
The reason is that the lightcone singularities of correlators have a richer structure than 
those of individual cross-channel conformal blocks or their finite sums. 
One may encounter divergent sums if the lightcone limit is taken before the summation over intermediate operators. 
For $SL(2,\mathbb R)$ blocks, this problem can be resolved by the resummation identities in \cite{Simmons-Duffin:2016wlq}, 
at least order by order.  
The asymptotic behaviour at large spin is associated with 
the emergent singularities, which cannot be produced by a finite sum of blocks. 
By adding and subtracting the corresponding identities, the remaining summation gives finite regular terms. 
Using the factorized decomposition \eqref{factorized-decomposition}, 
we can extend this technique to the resummations of full conformal blocks. 
The leading divergences are regularized by a few $SL(2,\mathbb R)$ resummation identities for one cross-ratio
multiplied by simple functions in the other cross-ratio \cite{Li:2019cwm}. 

The leading divergences are determined by the large spin behaviour, 
so we only need the low order terms of conformal blocks in the large-spin expansion. 
In the small $(1-\bar z)$ expansion, 
the basis functions $f_{k,m}$ are of order $\ell^{-2k}$ compared to $f_{0,0}$, 
so we can use the low order terms in \eqref{factorized-decomposition} to derive the leading terms in large spin. 
We have used the fact that the $\bar z$-independent part, i.e. $C_{k,m,n}$ and $g_{k,n}$, is at most of order $\ell^0$. 

For fixed twist, the higher order terms in the small $z$ expansion of $g_{k,n}(z)$ are suppressed by large spin, 
which simplifies the $n$-summation. 
As an example, let us consider the $k=0$ case. 
Here we also have $m=0$ due to $m\le k$. 
The large-spin expansion reads
\be
&&(1-z)^{a-b}\sum_{n=0}^\infty C_{0,0,n}\,g_{0,n}(z)
\sim
\sum_{p=0}^\infty\sum_{q=-p}^p
\frac{D_{0}^{p,q}}{J^{2p}}
\bigg(\frac{z}{1-z}\bigg)^{\frac{d-2} 2+p}\,
k_{\g+2q}(z)\,,
\label{small-zb-large-spin}
\ee
where the basis functions are given by $SL(2,\mathbb R)$ block $k_{\g}(z)$ with shifted parameters. 
The low order coefficients are
\be
D_{0}^{0,0}=1\,,\quad
D_{0}^{1,-1}=0\,,\quad
D_{0}^{1,0}=-\frac{\frac {d-2}2\frac {d-4}2\prod_{\a=-a,b}\big(\g/2+\a\big)}{\g}\,,
\label{D000}
\ee
\be
D_{0}^{1,1}=\frac{\frac {d-2}2\frac {d-4}2\prod_{\a=\pm a, \pm b}\big(\g/2+\a\big)}{\g^2(\g+1)}\,.
\label{D011}
\ee
We notice that the coefficients $D_0^{p,-p}$ always vanish when $p>0$. 
The large-spin expansion is naturally organized by 
\be
J^2=\frac 1 4\,\b(\b-2)\,,
\ee
which is the eigenvalue of the $SL(2,\mathbb R)$ Casimir operator associated with the direct-channel lightcone. 
Since the large-spin expansion is based on the small $z$ expansion, it is more accurate at small $z$, and less near $z=1$. 
We examine the large-spin expansion \eqref{small-zb-large-spin} in some concrete cases with $d=3$, $\g=1/3$ and $a=b=0$. 
The zeroth order approximation already reaches the $10^{-2}$ precision, i.e. the relative error is about or smaller than $10^{-2}$, 
when $\ell\geq 2$ and $1-z>0.02$.  
In the case of first order approximation, the precision reaches $10^{-3}$ for $\ell\geq 6$, $1-z>0.04$. 
For second order approximation, the precision goes to $10^{-4}$ if $\ell\geq 10$ and $1-z>0.04$. 
At larger spin, the good precision extends to almost the full range $0\leq z<1$ as expected.  

In general, the large-spin expansion of a $k$-th order term reads
\be
&&v^{a-b}\sum_{m=0}^k\sum_{n=0}^\infty C_{k,m,n}\,f_{k,m}(1-\bar z)\,g_{k,n}(z)
\nn&\sim&
\sum_{p=k}^\infty\sum_{q=-(p-k)}^{(p-k)}
\frac{D_{k}^{p,q}}{J^{2p}}\,
\Big(\frac z{1-z}\Big)^{\frac{d-2} 2+p}\,k_{\g+2q}(z)\,
(1-\bar z)^{a-b}\,f_{0,0}(1-\bar z)\,.
\ee
Note that $D_{k}^{p,q}$ with $k>0$ can contain derivatives. 
For example, the lowest order coefficient for $k=1$ is given by
\be
D_1^{1,0}&=&\frac {d-2}2\Big(\frac{d-4}2+\pa_{\log(1-\bar z)}\Big)\, \Big(-a+b+\pa_{\log(1-\bar z)}\Big)\,,
\label{D110}
\ee
where the derivatives with respect to $\log(1-\bar z)$ encode the differences between $f_{0,0}(1-\bar z)$ and $f_{k,m}(1-\bar z)$ 
near $\bar z=1$ at large spin. 
Combining the two parts of the factorized lightcone expansion in \eqref{factorized-decomposition}, we obtain
\be
G^{(d,a,b)}_{\t,\ell}(z,\bar z)\sim\frac{\G(\b/2)^2}{\G(\b)}\sum_{k=0}^\infty\sum_{p=k}^\infty\sum_{q=-(p-k)}^{p-k}\frac{ D_{k}^{p,q}}{J^{2p}}\,
\Big(\frac z{1-z}\Big)^{\frac{d-2} 2+p}\,k_{\g+2q}(z)\,k_{\b}(\bar z)\,,
\label{large-spin-expansion-general}
\ee
where the $\bar z$ dependence is encoded in terms of $SL(2,R)$ blocks as well. 
More explicitly, the low order terms in the large-spin expansion are
\be
G^{(d,a,b)}_{\t,\ell}(z,\bar z)&\sim&\frac{\G(\b/2)^2}{\G(\b)}\Big(\frac z{1-z}\Big)^{\frac{d-2} 2}
\bigg[\Big(1+\frac {D_{0}^{1,0}+D_{1}^{1,0}}{J^2}\frac z{1-z}+\dots\Big)k_{\g}(z)\,
\nn&&\qquad\qquad\qquad\qquad\qquad
+\Big(\frac {D_{0}^{1,1}}{J^2} \frac z{1-z}+\dots\Big)k_{\g+2}(z)
+\dots
\bigg]k_{\b}(\bar z)\,.
\label{large-spin-expansion-explicit}
\ee
Note that $D_0^{1,0}+D_1^{1,0}$ also generates the correct coefficients for the other part with leading power law $v^0$, 
which are associated with the $v^{a-b}$ coefficients by $a\leftrightarrow b$. 
Since the large-spin expansion is based on the lightcone expansion, 
one should be more careful about the nonperturbative effects 
if the near $z=1$ or $\bar z=0$ contributions are important. 
We have used $\sim$ to indicate the asymptotic nature. 

Using the simplified factorization at large spin, we can readily derive twist conformal blocks \cite{Alday:2016njk} , 
which are defined as infinite sums of conformal blocks with identical twist. 
The direct-channel lightcone limit plays the role of a boundary condition. 
If the small $z$ limit of a twist conformal block is given by a power law $(1-\bar z)^p$, 
then we have
\be
\sum_{\ell}\,P_{\t,\ell}\,G_{\t,\ell}(z,\bar z)
&\sim&
y^{\frac{d-2} 2}k_{\g}(z)\,(1-\bar z)^p
\nn&&
+\,y^{\frac{d-2} 2+1}
\frac{\big(D_{0}^{1,0}+D_{1}^{1,0}\big)k_{\g}(z)\,
+D_{0}^{1,1}k_{\g+2}(z)}
{(p+1)(p+1-a+b)}
{(1-\bar z)^{p+1}}\nn&&
+\,\mathcal O\big((1-\bar z)^{p+2}\big)\,,
\label{TCB-subleading}
\ee
where only the emergent power laws have been written explicitly 
and we have used the natural variable $y=\frac z{1-z}$. 
The large-spin expansion parameter $J^{-2}$ corresponds to 
an inverse action of the $SL(2,\mathbb R)$ Casimir operator on $(1-\bar z)^p$, 
which increases the power to $p+1$ and generates the denominator $(p+1)(p+1-a+b)$. 
The leading large-spin expansion coefficients $D_{k}^{p,q}$ can be found in \eqref{D000}, \eqref{D011}, \eqref{D110}. 
In particular, the derivatives become $p+1$, so we have
\be
D_1^{1,0}\rightarrow
\frac {d-2}2\Big(\frac{d-4}2+p+1\Big)\, (-a+b+p+1)\,.
\ee
As shown in \cite{Alday:2016njk}, the twist conformal blocks also satisfy a differential equation. 
By combining the quadratic and quartic Casimir equations, 
one can remove the spin dependence, and derive a fourth order differential equation. 
In principle, the coefficients of the series solution can be computed order by order using this quartic differential equation, 
but the complexity grows rapidly at high orders. 
As a consistency check,  
we verify that the series solution of the quartic differential equation resums to \eqref{TCB-subleading}. 
For $a=b=0$, the leading term, i.e.  the first line of \eqref{TCB-subleading}, was obtained in \cite{Alday:2016njk}. 
Here we give the general $(a,b)$ results to subleading order, 
and it is straightforward to go to higher orders.  
Our approach seems to be more efficient 
as the results are directly expressed in terms of the simple $SL(2,\mathbb R)$ blocks, 
which is not obvious if one only knows the low order series coefficients. 

We have explained how to determine the lightcone singularities from 
a resummation of conform blocks with identical twist 
based on the small $z$ limit. 
It is straightforward to generalize the method to the case of Regge trajectories. 
We should expand the twists around the large spin limit. 
More explicitly, the resummation of the large-spin tail of a Regge trajectory can be computed as 
\be
\sum_{\ell}f_{\ell}^2\,G_{\t,\ell}(z,\bar z)
&=&
\sum_{\ell}\frac{\pa\b}{\pa\bar\b}\,
\l_{\b}^2\,
\sum_{k,p,q} \frac{D_k^{p,q}(\g)}{(\frac \b 2(\frac \b 2-1))^{p}}\,
y^{\frac{d-2} 2+p}\,
k_{\g+2q}(z)\,k_{\b}(\bar z)
\nn&=& \sum_{m=0}^\infty\frac{(\pa_{\bar\t})^{m}}{m!}
\sum_{k,p,q}y^{\frac{d-2} 2+p}
k_{\bar \g+2q}(z)D_k^{p,q}(\bar\g)
\sum_{\ell}
\frac{(\d\t_{\bar \b})^{m}\,\l_{{\bar \b}}^2}{(\frac{\bar \b} 2(\frac {\bar \b} 2-1))^{p}}\,
k_{{\bar \b}}(\bar z)\,,\qquad
\ee
where $\bar \b=\bar \t+2\ell$, $\bar \g=\bar \t-d+2$ and $\d\t_\b=\t_\b-\bar \t$. 
Note that we have isolated the Jacobian ${\pa\b}/{\pa\bar\b}$ in the first line, 
which leads to $\bar \t$-derivatives and $\d\t_{\bar \b}$ insertions in the second line. 
\footnote{The small $z$ limit reduces to the case of $SL(2,\mathbb R)$ blocks discussed in \cite{Simmons-Duffin:2016wlq}
\be
\sum_{\ell}f_{\ell}^2\,k_{\b}(\bar z)
&=&
\sum_{\ell}
\frac{\pa\b}{\pa\bar\b}\,
\l^2_{\b}\,k_{\b}(\bar z)
=
\sum_{m=0}^\infty \frac {(\pa_{\bar \t})^m} {m!}\,
\sum_{\ell}
(\d\t_{\bar \b})^m\,
\l^2_{\bar\b}\,k_{\bar \b}(\bar z)
\,,
\ee
which contains the main features. }
Then we can perform the resummations of large-spin tails as in the case of identical twist. 
If the finite spin data deviates from the large spin behaviour significantly, 
we should not expand the twist around the large spin limit. 
We can still use \eqref{large-spin-expansion-general} for relatively large conformal spin. 
At low conformal  spin, we should use the exact formula \eqref{factorized-decomposition-tilde}. 

For the lightcone bootstrap, the Lorentzian inversion of the resummed Regge trajectories 
in the cross-channel involves the basic $\bar z$-integral
\be
&&\int_0^1 \frac{d\bar z}{\bar z^2}\,
k^{(-a,-b)}_\b(\bar z)\,
\frac{\bar z^{(\D_1+\D_2)/2}}{(1-\bar z)^{(\D_2+\D_3)/2}}\,
\frac{(1-\bar z)^{p}}{\bar z^{p}}\,k^{(a',b')}_{\g'}(1-\bar z)
\nn&=&
\G(\b)\,\G(\g')\,\G\Big(\frac{\b+\D_1+\D_2}2-p-1\Big)\,
\G\Big(\frac{\b+\D_3+\D_4}2-p-1\Big)
\nn&&\times\,
\G\Big(\frac{\g'-\D_1-\D_4}2+p+1\Big)\,
\G\Big(\frac{\g'-\D_2-\D_3}2+p+1\Big)
\nn&&\times\,
\psi\Big(\frac{\b+2\g'-\D_1+\D_2}2-1;\,\frac{\b-\D_1+\D_2}2,
\frac{\g'-\D_1+\D_4}2,
\frac{\g'+\D_2-\D_3}2,
\nn&&\qquad
\frac{\g'-\D_1-\D_4}2+p+1,
\frac{\g'+\D_2+\D_3}2-p-1
\Big)\,,
\label{kk-integral}
\ee
which can be viewed as a special case of \eqref{g-integral}. 

\section{Conclusions}
We have presented a new complete formula \eqref{factorized-decomposition} 
for the lightcone expansion of 4-point scalar conformal blocks in factorized form. 
Although it applies to arbitrary intermediate spin and general spacetime dimensions, 
the factorized decomposition takes a remarkably simple form. 
For physical spin $\ell$, an equivalent formula has been given in \eqref{factorized-decomposition-tilde}, 
where two summations terminate at order $\ell$. 
As an application of the general formula \eqref{factorized-decomposition}, 
we discuss the large-spin expansion \eqref{large-spin-expansion-general}, 
which can be systematically derived from \eqref{factorized-decomposition}. 
We further explain how to resum the large-spin tails of Regge trajectories using the factorization structure. 
In \eqref{g-integral} and \eqref{kk-integral}, the basic integrals for the general Lorentzian inversion have been carried out, 
and the results are given by Wilson functions. 
In this work, we consider generic external scalar operators. 
The factorization structure should extends to the cases of external spinning operators, 
and the especially cases are those involving external conserved currents \cite{Li:2015itl,Hofman:2016awc}.

As our results apply to general spacetime dimensions, 
it would be interesting to consider the analytic bootstrap of the Wilson-Fisher fixed point in fractional dimensions. 
This was first studied by the numerical bootstrap in \cite{El-Showk:2013nia}.  
A subtlety about non-integer dimensions is that there exists negative-norm states \cite{Hogervorst:2015akt}. 
\footnote{If the non-unitary effects are small, 
one can obtain reasonable results from the numerical method based on unitarity, 
up to certain precision. 
This is similar to the case of $O(N)$ models with non-integer $N$ \cite{Shimada:2015gda}, 
whose non-unitary nature was recently discussed in the language of Deligne categories \cite{Binder:2019zqc}. 
It would also be interesting to study them using the truncation methods \cite{Gliozzi:2013ysa, Li:2017ukc}. 
The alternative approach does not use unitarity, so is more suitable for many important problems in statistical physics 
\cite{Gliozzi:2014jsa,Gliozzi:2015qsa,Hogervorst:2016itc,Gliozzi:2016cmg,Hikami:2017hwv,Hikami:2017sbg,Hikami:2018mrf,LeClair:2018edq,Hikami:2018qpz,Kaviraj:2019tbg}. 
To obtain stronger results, one may need to combine them with the analytic methods, 
in order to reduce the number of independent parameters,  
as proposed in \cite{Simmons-Duffin:2016wlq}.}
The analytic bootstrap can provide a complementary approach to 
study the smooth deformation from Gaussian models to strongly interacting models, 
without using unitarity or an asymptotic expansion in $\e=d_0-d$. 

In \cite{Ising-revisit}, we revisit 
the mixing problem of the 3d Ising model \cite{Simmons-Duffin:2016wlq,Caron-Huot:2020ouj} 
using the new results presented in this paper. 
The resolution of the mixing problem gives rise to 
the repulsion of near-degenerate Regge trajectories and vanishing OPE coefficients at low spin. 
The lower trajectory becomes non-unitary below the decoupling spin, so is in conflict with unitarity. 
We believe that this is the analytic origin of the 3d Ising kink \cite{ElShowk:2012ht,El-Showk:2014dwa}. 
In fact, the kink phenomena are ubiquitous in the numerical bootstrap studies. 
It is straightforward to generalize the above analytic interpretation. 

Further numerical bootstrap studies have led to the beautiful discoveries of precision islands 
\cite{Kos:2014bka,Kos:2015mba,Kos:2016ysd,Rong:2018okz,Atanasov:2018kqw,Chester:2019ifh,Chester:2020iyt}. 
To obtain an isolated region, one needs at least three cuts in a two-parameter space, 
where one cut separates a non-unitary region from a potentially unitary region. 
\footnote{It is also important to impose gap assumptions about the scaling dimensions. 
The particularly well-motivated ones are associated with the number of relevant operators, 
which implies that the other operators should have scaling dimensions greater than $d$. 
} 
It was shown in \cite{Kos:2014bka} that an isolated region can be obtained by considering the mixed correlator. 
Previously, two sides of the 3d Ising kink have been shown to be related to the decoupling of subleading operators of spin-0 and spin-2 \cite{ElShowk:2012ht,El-Showk:2014dwa}. 
\footnote{
For the spin-0 decoupling, one may need to use analytic continuation. 
For the Wilson-Fisher CFT, the analytic continuation to the lowest $\mathbb Z_2$-even scalar was shown in \cite{Alday:2017zzv}, 
based on the perturbative expansion in $\e=4-d$. 
For the 3d Ising and O(2) CFTs, this phenomenon has also been studied in \cite{Caron-Huot:2020ouj,Liu:2020tpf}. 
(See \cite{Caron-Huot:2020ouj} for the relation to asymptotic transparency.) 
The fact that the continuation gives the correct results, i.e. the low-lying operators also lie on the Regge trajectories,  
implies that physical correlators have exceptionally good behaviour beyond the Euclidean regime, 
and hence nontrivial constraints on the full spectrum, 
which is similar to the relation between a bounded Regge limit and analyticity in spin above spin-1. 
}
The third side should be related to the leading $\mathbb Z_2$-odd trajectory,  
where the continuation to spin-0 and the spin-1 decoupling were recently discussed in \cite{Caron-Huot:2020ouj}. 
To derive precision islands, one needs to be sensitive to more non-unitary sectors, 
which are associated with subleading Regge trajectories. 
A larger system of crossing equations can capture more trajectories and 
help to efficiently resolve the subleading mixing problems. 
The importance of considering more crossing equations has been noticed in the numerical studies of precision islands \cite{Chester:2019ifh,Chester:2020iyt}. 
\footnote{
In 2d, the Regge trajectories have simpler structure due to the Virasoro symmetry. 
One side of the 2d Ising kink \cite{Rychkov:2009ij} has an analytic interpretation  
as a one-parameter family of crossing solutions \cite{Liendo:2012hy}. 
If we assume the twists of the low-lying trajectories are partly constrained, 
we can ``define" the 2d Lee-Yang and Ising CFTs 
as the decoupling points, 
where the crossing constraint implies that the OPE coefficient of a low-lying spin-2 operator vanishes \cite{Li:2017agi}. 
Perhaps we can extend this definition to higher dimensions using the powerful analyticity in spin. 
The decoupling of more operators from the null state conditions of Virasoro multiplets should be related to the decoupling in higher Regge trajectories.  
}
We plan to apply our new results to the analytic study of precision islands. 

\begin{acknowledgments}
I would like to thank Simon Caron-Huot, Shinobu Hikami, Yue-Zhou Li, David Meltzer, Junchen Rong and Ning Su  
for inspiring discussions or correspondence. 
This work was supported by Okinawa Institute of Science and Technology Graduate University (OIST) 
and JSPS Grant-in-Aid for Early-Career Scientists (KAKENHI No. 19K14621).

\end{acknowledgments}

\appendix
\renewcommand{\theequation}{\thesection.\arabic{equation}}
\section{Factorized decomposition of Appell's $F_4$ function}
The integer-balanced $F_4$ functions admit finite decompositions.
For $B_1+B_2=1+A_1+A_2$, i.e. $1$-balanced, the $F_4$ function factorizes into a product of two ${}_2F_1$ functions \cite{Bailey}
\be
F_4\bigg[
\begin{matrix}
A_1, A_2\\
B_1,B_2
\end{matrix}; x(1-y),y(1-x)
\bigg]=
{}_2F_1\bigg[
\begin{matrix} 
A_1,A_2 \\
B_1
\end{matrix}
;\,x\bigg]\,
{}_2F_1\bigg[
\begin{matrix} 
A_1,A_2 \\
B_2
\end{matrix}
;\,y\bigg]\,.
\label{F4-1-balanced}
\ee
For $B_1+B_2=A_1+A_2$, i.e. $0$-balanced, there exists a 2-term decomposition:
\be
&&(1-x-y)\,F_4\bigg[
\begin{matrix}
A_1, A_2\\
B_1,B_2
\end{matrix}; x(1-y),y(1-x)
\bigg]\nn&=&
{}_2F_1\bigg[
\begin{matrix} 
A_1-1,A_2-1 \\
B_1-1
\end{matrix}
;\,x\bigg]\,(1-y)\,{}_2F_1\bigg[
\begin{matrix} 
A_1,A_2 \\
B_2
\end{matrix}
;\,y\bigg]
\nn&&+\frac{(A_1-B_1)(A_2-B_1)}{B_1(B_1-1)}
\,x\,{}_2F_1\bigg[
\begin{matrix} 
A_1,A_2 \\
B_1+1
\end{matrix}
;\,x\bigg]\,{}_2F_1\bigg[
\begin{matrix} 
A_1-1,A_2-1 \\
B_2
\end{matrix}
;\,y\bigg]\,.
\label{F4-0-balanced}
\ee
For a generic $F_4$ function, the factorized decomposition found by Burchnall and Chaundy 
reads \cite{Burchnall-Chaundy} 
\be
F_4\bigg[
\begin{matrix}
A_1, A_2\\
B_1,B_2
\end{matrix}; x(1-y),y(1-x)
\bigg]
=
\sum_{k=0}^\infty
\frac{(1+A_1+A_2-B_1-B_2)_k}{k!}
\frac{(A_1)_k\,(A_2)_k}{(B_1)_k\,(B_2)_k}\,
\nn\times\,
x^k\,{}_2F_1\bigg[
\begin{matrix} 
A_1+k,A_2+k \\
B_1+k
\end{matrix}
;\,x\bigg]\,
y^k\,{}_2F_1\bigg[
\begin{matrix} 
A_1+k,A_2+k \\
B_2+k
\end{matrix}
;\,y\bigg]\,.\quad
\label{F4-decomposition}
\ee
In terms of derivatives, the above decomposition can be written as a product of three hypergeometric series
\be
F_4\bigg[
\begin{matrix}
A_1, A_2\\
B_1,B_2
\end{matrix}; x(1-y),y(1-x)
\bigg]
&=&
{}_3F_2\bigg[
\begin{matrix} 
1+A_1+A_2-B_1-B_2\,, -\pa_{\log x}, -\pa_{\log y}\\
A_1\,, A_2
\end{matrix}
;\,1\bigg]
\nn&&\times\,
{}_2F_1\bigg[
\begin{matrix} 
A_1,A_2 \\
B_1
\end{matrix}
;\,x\bigg]\,
{}_2F_1\bigg[
\begin{matrix} 
A_1,A_2 \\
B_2
\end{matrix}
;\,y\bigg]\,,
\label{F4-derivative}
\ee
which can be viewed as a generalization of \eqref{F4-1-balanced}. 

\section{More on the cross-channel lightcone limit}
The Kamp\'e de F\'eriet function $F^{0,2,2}_{0,2,1}$ is defined as
\be
&&F^{0,2,2}_{0,2,1}
\bigg[\Big|
\begin{matrix}
A_1, A_2\\
B_1,B_2
\end{matrix}
\Big|
\begin{matrix}
A_3, A_4\\
B_3
\end{matrix}
\Big|
x, y
\bigg]
=\sum_{m,n=0}^\infty\,
 \frac{(A_1)_n\,(A_2)_n\,}{(B_1)_n\,(B_2)_n}\,
  \frac{(A_3)_{m+n}\,(A_4)_{m+n}}{(B_3)_{m+n}}\,
  \frac{x^m\,y^n}{m!\,n!}\,.\quad
 \ee
The cross-channel lightcone limit of conformal blocks is associated with the case of $y=-x$, 
which can be expressed in a product form using derivatives
\be
&&F^{0,2,2}_{0,2,1}
\bigg[\Big|
\begin{matrix}
A_1, A_2\\
B_1,B_2
\end{matrix}
\Big|
\begin{matrix}
A_3, A_4\\
B_3
\end{matrix}
\Big|
x, -x
\bigg]
={}_3F_2\bigg[
\begin{matrix} 
A_1, A_2, -\pa_{\log x}\\
B_1\,, B_2
\end{matrix}
;\,1\bigg]\,
{}_2F_1\bigg[
\begin{matrix} 
A_3, A_4\\
B_3
\end{matrix}
;\,x\bigg]\,.
 \ee
For even $d$, the cross-channel channel lightcone limit of a conformal block reduces to a finite sum of ${}_2F_1$ functions. 
In \eqref{cross-lightcone}, the $n$-summation terminates for physical $\ell$ in general dimensions. 
Based on the even-dimensional results, we find an alternative expression such that the even-$d$ termination becomes manifest. 
The Kamp\'e de F\'eriet function in \eqref{cross-lightcone} can be written as
\be
&&
F^{0,2,2}_{0,2,1}
  \bigg[
\Big|
\begin{matrix}
 -\ell,\, 3-d-\ell \\
2- d/ 2 -\ell\,, \g
\end{matrix}\,
\Big|
\begin{matrix}
  \g/2-a , \, \g/ 2+b\\
 \b/2+\g/ 2
\end{matrix}\,\Big| 
z,-z
\bigg]
\nn&=&
F^{0,2,2}_{0,2,1}
  \bigg[
\Big|
\begin{matrix}
(d-2)/2,\, -(d-4)/2 \\
2- d/ 2 -\ell,\,\b/2+\g/2
\end{matrix}\,
\Big|
\begin{matrix}
  \g/2-a , \, \g/ 2+b\\
 \g
\end{matrix}\,\Big| 
z,-z
\bigg]
\nn&=&
\sum_{n=0}^\infty 
\frac{(d/2-1-n)_{2n}\,(\g/2-a)_n\,(\g/2+b)_n}{(2-d/2-\ell)_n\,(\b/2+\g/2)_n\, (\g)_n}
\,\frac {z^n}{n!}\,
{}_2F_1\bigg[
\begin{matrix} 
\g/2-a+n,\g/2+b+n \\
\g+n
\end{matrix}
;\,z\bigg]\,,\quad
\label{t-lightcone-even-d}
\ee
where two parameters in the bottom row are interchanged, and two parameters in the top row are modified. 
For even $d$, the n-summation terminates due to $(d/2-1-n)_{2n}$, 
so there are at most $(1+|d-3|)/2$ independent ${}_2F_1$ functions in the cross-channel lightcone limit. 

\bibliographystyle{JHEP}

\end{document}